\documentclass[twocolumn]{aastex631}

\usepackage{color}
\usepackage{amsmath}
\usepackage[T1]{fontenc}

\newcommand{\columbia}{Department of Astronomy, Columbia University, 550 West 120th Street, New York, NY 10027, USA}

\newcommand{\lafayette}{Department of Physics, Lafayette College, 730 High St, Easton, PA 18042, USA}

\shorttitle{The Spectroscopic UV Evolution of K Stars}
\shortauthors{Richey-Yowell et al. (2022)}

\begin{document}
\title{HAZMAT. VIII. A Spectroscopic Analysis of the Ultraviolet Evolution of K Stars: Additional Evidence for K Dwarf Rotational Stalling in the First Gigayear}

\author[0000-0003-1290-3621]{Tyler Richey-Yowell}
\affil{School of Earth and Space Exploration, Arizona State University, Tempe, AZ 85287, USA}
\email{try@asu.edu}

\author[0000-0002-7260-5821]{Evgenya L. Shkolnik}
\affil{School of Earth and Space Exploration, Arizona State University, Tempe, AZ 85287, USA}

\author[0000-0001-5646-6668]{R. O. Parke Loyd}
\affil{School of Earth and Space Exploration, Arizona State University, Tempe, AZ 85287, USA}
\affil{Eureka Scientific, 2452 Delmer Street Suite 100, Oakland, CA, 94602-3017, USA}

\author[0000-0003-0711-7992]{James A. G. Jackman}
\affil{School of Earth and Space Exploration, Arizona State University, Tempe, AZ 85287, USA}

\author[0000-0002-6294-5937]{Adam C. Schneider}
\affil{U.S. Naval Observatory, 10391 W Naval Observatory Rd, Flagstaff, AZ 86001, USA}
\affil{Department of Physics and Astronomy, George Mason University, MS3F3, 4400 University Drive, Fairfax, VA 22030, USA}

\author[0000-0001-7077-3664]{Marcel A.~Ag\"{u}eros}
\affil{\columbia}

\author[0000-0002-7129-3002]{Travis Barman}
\affil{Lunar and Planetary Laboratory, University of Arizona, Tucson, AZ 85721, USA}

\author[0000-0002-1386-1710]{Victoria S. Meadows}
\affil{NASA Astrobiology Institute, Virtual Planetary Laboratory, University of Washington, Seattle, WA 85215, USA}

\author{Rose Gibson}
\affil{\columbia}

\author[0000-0001-7371-2832]{Stephanie T.\ Douglas}
\affiliation{\lafayette}

\begin{abstract}

Efforts to discover and characterize habitable zone planets have primarily focused on Sun-like stars and M dwarfs. K stars, however, provide an appealing compromise between these two alternatives that has been relatively unexplored. Understanding the ultraviolet (UV) environment around such stars is critical to our understanding of their planets, as the UV can drastically alter the photochemistry of a planet's atmosphere. Here we present near-UV and far-UV \textit{Hubble Space Telescope}'s Cosmic Origins Spectrograph observations of 39 K stars at three distinct ages: 40 Myr, 650 Myr, and $\approx$5 Gyr. We find that the K star (0.6 -- 0.8 M$_{\odot}$) UV flux remains constant beyond 650 Myr before falling off by an order of magnitude by field age. This is distinct from early M stars (0.3 -- 0.6 M$_{\odot}$), which begin to decline after only a few hundred Myr. However, the rotation-UV activity relation for K stars is nearly identical to that of early M stars. These results may be a consequence of the spin-down stalling effect recently reported for K dwarfs, in which the spin-down of K stars halts for over a Gyr when their rotation periods reach $\approx$10 d, rather than the continuous spin down that G stars experience. These results imply that exoplanets orbiting K dwarfs may experience a stronger UV environment than thought, weakening the case for K stars as hosts of potential ``super-habitable'' planets.

\end{abstract}


\submitjournal{ApJ}
\accepted{March 11, 2022}
\section{Introduction}\label{sec:intro}

While often overlooked in favor of M and G stars, K stars have been gaining traction as potential ``Goldilocks'' hosts for habitable planets. Many factors go into determining the potential habitability of a planet from the stellar perspective, e.g., frequency and lifetime of the star, contraction time onto the main sequence, the size and longevity of the habitable zone (HZ), ultraviolet (UV) and X-ray emission, stellar variability, and the potential for tidal locking \citep[e.g.,][]{Meadows2018}. Compared to solar-type stars, K stars are more abundant, maintain longer main-sequence lifetimes, and their planets are more suitable to observations based on the mass and radius ratios of the planet to star. While M stars also excel in these regards, they have other potentially life-destroying disadvantages compared to K dwarfs. For example, superflares (flares more energetic than the largest recorded solar events, i.e., \citealt{Carrington1859, Tsurutani2003}) appear to occur daily on young M stars \citep[][]{Loyd2018hazmat}. Yet the probability of finding a habitable planet around solar-type stars decreases significantly due to the stars’ shorter lifetimes, the lower frequency of the stars, and more importantly, our difficulty in detecting and characterizing their planets. HZ planets around old solar-type stars are too faint to detect directly, have smaller RV amplitudes, and transit only once per year. 

The UV radiation incident on a planet (both through quiescent emission and flaring) is a key factor in assessing the habitability the planet, as the UV has the potential to alter the chemistry of planetary atmospheres by ionizing and photo-dissociating both the most important molecules for the development of life and signatures that would be indicative of life, with the potential for complete erosion of the atmosphere \citep{Airapetian2017}. Therefore, in order to gain an understanding of the atmospheric environments of planets around any type of star today, it is essential to characterize the distribution of UV properties and how these properties evolve with age. 

\subsection{The Effect of UV Radiation on Planetary Atmospheres}

The near-UV (NUV, 1700-3200 \AA), far-UV (FUV, 912-1700 \AA), and extreme-UV (EUV, 100-912 \AA) each affect planetary atmospheres differently. The NUV and FUV radiation will photo-dissociate the molecules in the atmosphere, including those most important for the development and detection of life (e.g. H$_2$O, O$_2$, OCS, and NH$_3$) \citep[e.g.][]{Lichtenegger2010, Segura2010, Hu2012}. The FUV to NUV flux ratio determines the lifetimes of biogenic species such as CH$_4$, N$_2$O, and CH$_3$Cl, allowing for a greater probability of detection with a greater FUV to NUV ratio \citep{Segura2005}, but will also increase abiotic oxygen and ozone, thus producing higher chances of false-positive biosignatures \citep[e.g.][]{Domagal-Goldman2014, tian14, harman15}. \citet{richey-yowell2019} measured this ratio to be up to 10 times smaller for K stars than for M stars, which have been measured photometrically to be between $\approx0.1 - 0.3$ \citep[][]{Schneider2018} and spectroscopically to be between $\approx0.5 - 3$ \citep[][]{france2013}, giving K stars an added advantage for the detection of real biosignatures. The EUV values are crucial since the EUV will heat and ionize the atmosphere, powering thermal atmospheric escape  \citep[][]{Lammer2003, Lammer2007, Koskinen2010}. However, the EUV flux from stars other than the Sun is predominantly absorbed by the interstellar medium. Limited data exist in the EUV and must therefore be estimated from empirical relationships, differential emission measure models, or stellar atmospheric models constrained by data at FUV, NUV, or X-ray wavelengths \citep[e.g.,][]{Pagano2009, Linsky2015, Fontenla2016, Peacock2019, Peacock2020, Tilipman2021}.

The photo-dissociation of the molecules in planetary atmospheres can also lead to increased production of hazes in reducing atmospheres of sub-Neptunes and Archean Earth (Earth 4.0 -- 2.5 Myr ago) analogs, thus drastically altering the spectrum and the detectability of biosignatures \citep[][]{Zerkle2012, Arney2017}. For future observations such as with the \textit{James Webb Space Telescope } (\textit{JWST}) that will focus on planetary spectra, hazes could limit the detectability of spectral features in the atmosphere \citep[e.g.,][]{Fauchez2019}. While no HZ planets around K dwarfs are expected to be seen by TESS due to orbital period constraints \citep[][]{barclay18}, $\sim$30 HZ K dwarf terrestrial planets are expected to be discovered in the next decade with ESA’s PLATO mission (PLATO Definition Study Report, 2017). 

\subsection{The K Dwarf Advantage}

Several studies have recently focused on K stars as optimal hosts for habitable planets. Both \citet{Cuntz2016} and \citet{Heller2014} found that early-type K stars may offer the most suitable conditions for life and may even be “super-habitable,” i.e., even more habitable than an Earth-sized planet around a G2V star. From a photochemistry perspective, studies have demonstrated that K6V stars ($\approx$0.7 M$_{\odot}$) may be the most suitable laboratories for producing detectable CH$_4$ and O$_2$ \citep{Arney2017, Arney2019}, while the increased O$_2$ of planets around these stars would permit complex life to develop much faster \citep{Lingam2017}.

\subsection{K Star Activity Evolution}
Using the Galaxy Evolution Explorer (GALEX) broadband UV photometry of K stars ranging in age from 10 Myr to $\approx$5 Gyr, \citet{richey-yowell2019} compared the photometric UV evolution of K stars to that of early- and late-type M stars \citep{Shkolnik2014, Schneider2018}. They reported that the intrinsic NUV radiation of K stars only decreased slightly (half an order of magnitude) between 10 Myr -- $\approx$5 Gyr (the average age for the field stars), and the FUV radiation decreased similarly from 10 -- 650 Myr before falling off more steeply (an order of magnitude) by 5 Gyr. This trend is different for late M stars, which are active longer because of their slower spin-down rate, and early M stars, which remain constant through 150 Myr before declining sharply (1.5 orders of magnitude) in both the NUV and FUV. However, GALEX observations only give us a narrow photometric window into the overall evolution of K star UV flux, as some of the strongest and thus most important emission features of these stars are outside of the GALEX bandpasses (e.g. C III, Si III, N V, C II, and Mg II). 

% If "omit" error, probably from vspace in the wrong place. Can't submit paper until that's fixed and no red errors. 
\begin{longrotatetable}
\begin{deluxetable}{l l c c c c c c c c c}
\centering
\tabletypesize{\scriptsize}
\tablecaption{\normalsize{Target information broken down by age.} \label{tab:target_info}}
\tablehead{
\colhead{Name} \vspace{-0.3cm} & \colhead{2MASS Name} & \colhead{RA} & \colhead{Dec.} & \colhead{SpT} & \colhead{Mass} & \colhead{Distance\tablenotemark{a}} & \colhead{Derived Age\tablenotemark{b}}& \colhead{Rotation Period\tablenotemark{c}} & \colhead{Ref.} & \colhead{Rossby }\\
\colhead{} & \colhead{} & \colhead{} & \colhead{} & \colhead{} & \colhead{[M$_{\odot}$]} & \colhead{[pc]} & \colhead{[Myr]} & \colhead{[days]} & \colhead{} & \colhead{Number}\\
\vspace{-0.5cm}
}
\startdata
\multicolumn{11}{l}{Tucana-Horologium Members (40 Myr) } \\																															
\hline																								
HIP 1910	&	J00240899-6211042	&	00 24 08.98	&	-62 11 04.29	&	K7.2	&	0.69	&	44.23	 $\pm$ 	1.07	&	40	 $\pm$ 5 &	1.751	 $\pm$ 	0.003	&	1	&	0.044	\\
CT Tuc	&	J00251465-6130483	&	00 25 14.66	&	-61 30 48.25	&	K5.7	&	0.72	&	44.16	 $\pm$ 	0.06	&	40	 $\pm$ 5 &	4.190	 $\pm$ 	0.419	&	1	&	0.121	\\
CD-46 644	&	J02105538-4603588	&	02 10 55.39	&	-46 03 58.65	&	K4.2	&	0.80	&	81.78	 $\pm$ 	0.61	&	40	 $\pm$ 5 &	1.116	 $\pm$ 	0.001	&	1	&	0.060	\\
CD-35 1167	&	J03190864-3507002	&	03 19 08.66	&	-35 07 00.29	&	K5.7	&	0.77	&	45.56	 $\pm$ 	0.07	&	40	 $\pm$ 5 &	8.5	 $\pm$ 	0.1	&	1	&	0.275	\\
CD-44 1173	&	J03315564-4359135	&	03 31 55.64	&	-43 59 13.54	&	M0	&	0.60	&	45.28	 $\pm$ 	0.07	&	40	 $\pm$ 5 &	2.94	 $\pm$ 	0.01	&	1	&	0.092	\\
TYC 8083-45-5	&	J04480066-5041255	&	04 48 00.66	&	-50 41 25.66	&	K5.7	&	0.72	&	59.60	 $\pm$ 	0.27	&	40	 $\pm$ 5 &	8.44	 $\pm$ 	0.05	&	1	&	0.224	\\
TYC 8098-41	&	J05341467-5145545	&	05 33 25.58	&	-51 17 13.10	&	K4.9	&	0.77	&	53.95	 $\pm$ 	2.67	&	40	 $\pm$ 5 &	5.22	 $\pm$ 	0.52	&	2	&	0.154	\\
HIP 107345	&	J21443012-6058389	&	21 44 30.12	&	-60 58 38.87	&	K7.2	&	0.69	&	46.36	 $\pm$ 	0.05	&	40	 $\pm$ 5 &	4.54	 $\pm$ 	0.02	&	1	&	0.120	\\
2MASS J23261069-7323498	&	J23261069-7323498	&	23 26 10.70	&	-73 23 49.89	&	K7.7	&	0.66	&	46.29	 $\pm$ 	0.06	&	40	 $\pm$ 5 &	0.57	 $\pm$ 	0.02	&	2	&	0.014	\\
UCAC3 13-31217 	&	J23585674-8339423	&	23 58 56.74	&	-83 39 42.3	&	K5.8	&	0.72	&	55.63	 $\pm$ 	0.06	&	40	 $\pm$ 5 &	...	 &	...	&	...	\\
\hline																															
\multicolumn{11}{l}{Hyades Members (650 Myr)} \\																															
\hline																															
2MASS J03510309+2354134	&	J03510309+2354134	&	03 51 03.11	&	+23 54 13.14	&	K6	&	0.79	&	40.69	 $\pm$ 	0.08	&	650	 $\pm$ 70 &	12.57	 $\pm$ 	1.26	&	3	&	0.603	\\
2MASS J03524101+2548159	&	J03524101+2548159	&	03 52 41.00	&	+25 48 15.97	&	K7	&	0.62	&	45.28	 $\pm$ 	0.12	&	650	 $\pm$ 70 &	14.66	 $\pm$ 	1.47	&	3	&	0.498	\\
2MASS J03550142+1229081	&	J03550142+1229081	&	03 55 01.44	&	+12 29 08.11	&	K0	&	0.63	&	45.92	 $\pm$ 	0.09	&	650	 $\pm$ 70 &	11.66	 $\pm$ 	1.17	&	3	&	0.628	\\
2MASS J04070122+1520062	&	J04070122+1520062	&	04 07 01.22	&	+15 20 06.10	&	K4	&	0.75	&	45.09	 $\pm$ 	0.24	&	650	 $\pm$ 70 &	14.03	 $\pm$ 	1.40	&	3	&	0.639	\\
2MASS J04081110+1652229	&	J04081110+1652229	&	04 08 11.09	&	+16 52 23.11	&	K7	&	0.62	&	40.10	 $\pm$ 	0.07	&	650	 $\pm$ 70 &	13.63	 $\pm$ 	1.36	&	3	&	0.398	\\
2MASS J04082667+1211304	&	J04082667+1211304	&	04 08 26.66	&	+12 11 30.64	&	K5	&	0.60	&	46.33	 $\pm$ 	0.15	&	650	 $\pm$ 70 &	12.96	 $\pm$ 	1.30	&	3	&	0.467	\\
2MASS J04172512+1901478	&	J04172512+1901478	&	04 17 25.15	&	+19 01 47.67	&	K5	&	0.71	&	47.86	 $\pm$ 	0.18	&	650	 $\pm$ 70 &	12.84	 $\pm$ 	1.28	&	3	&	0.557	\\
2MASS J04240740+2207079	&	J04240740+2207079	&	04 24 07.42	&	+22 07 07.92	&	K5	&	0.71	&	45.78	 $\pm$ 	0.12	&	650	 $\pm$ 70 &	13.01	 $\pm$ 	1.30	&	3	&	0.477	\\
2MASS J04293897+2252579	&	J04293897+2252579	&	04 29 38.99	&	+22 52 57.79	&	K5.5	&	0.68	&	59.35	 $\pm$ 	0.19	&	650	 $\pm$ 70 &	14.95	 $\pm$ 	1.50	&	3	&	0.671	\\
V925 Tau	&	J04303819+2254289	&	04 30 38.19	&	+22 54 28.83	&	M3	&	0.66	&	50.93	 $\pm$ 	0.26	&	650	 $\pm$ 70 &	2.50	 $\pm$ 	0.25	&	3	&	0.040	\\
2MASSJ04333716+2109030	&	J04333716+2109030	&	04 33 37.18	&	+21 09 03.06	&	K7	&	0.62	&	43.80	 $\pm$ 	0.09	&	650	 $\pm$ 70 &	13.59	 $\pm$ 	1.36	&	3	&	0.543	\\
2MASS J04334192+1900504	&	J04334192+1900504	&	04 33 41.92	&	+19 00 50.52	&	K7	&	0.62	&	47.80	 $\pm$ 	0.16	&	650	 $\pm$ 70 &	12.64	 $\pm$ 	1.26	&	3	&	0.542	\\
STKM 1-495\tablenotemark{d}	&	J04350255+0839304	&	04 35 02.55	&	+08 39 30.55	&	M1	&	0.79	&	59.34	 $\pm$ 	2.64	&	650	 $\pm$ 70 &	11.76	 $\pm$ 	1.18	&	4	&	0.284	\\
UCAC4 518-008871\tablenotemark{d}	&	J04451961+1334274	&	04 45 19.60	&	+13 34 27.34	&	K6	&	0.66	&	51.71	 $\pm$ 	0.71	&	650	 $\pm$ 70 &	2.22	 $\pm$ 	0.2203	&	3	&	0.035	\\
2MASS J04470892+2052564	&	J04470892+2052564	&	04 47 08.93	&	+20 52 56.32	&	K4	&	0.75	&	42.30	 $\pm$ 	0.09	&	650	 $\pm$ 70 &	10.54	 $\pm$ 	1.05	&	3	&	0.595	\\
2MASS J04472618+2303032	&	J04472618+2303032	&	04 47 26.18	&	+23 03 03.32	&	K4	&	0.75	&	55.09	 $\pm$ 	0.17	&	650	 $\pm$ 70 &	11.79	 $\pm$ 	1.18	&	3	&	0.579	\\
2MASS J04514917+1716255	&	J04514917+1716255	&	04 51 49.18	&	+17 16 25.43	&	K5	&	0.71	&	54.47	 $\pm$ 	0.15	&	650	 $\pm$ 70 &	13.01	 $\pm$ 	1.30	&	3	&	0.515	\\
\hline																															
\multicolumn{11}{l}{Field Stars ($\approx$5 Gyr)} \\																															
\hline																															
CD-23 1056 	&	J02464286-2305119	&	02 46 42.88	&	-23 05 11.80	&	K5	&	0.70	&	23.39	 $\pm$ 	0.02	&	1436	 $^{+	5981	 }_{-	379	}$ &	19	 $\pm$ 	2	&	5	&	0.582	 	\\
BD+16 502	&	J03435253+1640198	&	03 43 52.56	&	+16 40 19.30	&	K6	&	0.65	&	17.21	 $\pm$ 	0.01	&	1428	 $^{+	1033	 }_{-	402	}$ &	21.81	 $\pm$ 	2.18	&	6	&	0.610		\\
HD 266611	&	J06570468+3045235	&	06 57 04.68	&	+30 45 23.37	&	K5	&	0.70	&	18.89	 $\pm$ 	0.04	&	5889	 $^{+	4947	 }_{-	4247	}$ &	 ... 	&	...	&	 ... 	\\
HD 85512	&	J09510700-4330097	&	09 51 07.05	&	-43 30 10.02	&	K6	&	0.65	&	11.28	 $\pm$ 	0.01	&	4766	 $^{+	5130	 }_{-	3520	}$ &	 ... 	&	...	&	 ... 	\\
BD+57 1274	&	J10314321+5706571	&	10 31 43.22	&	+57 06 57.10	&	K5	&	0.70	&	17.56	 $\pm$ 	0.01	&	6219	 $^{+	4884	 }_{-	4311	}$ &	 ... 	&	...	&	 ... 	\\
HD 99492	&	J11264627+0300229	&	11 26 46.27	&	+03 00 22.75	&	K3	&	0.79	&	18.21	 $\pm$ 	0.06	&	8774	 $^{+	3069	 }_{-	3766	}$ &	 ... 	&	...	&	 ... 	\\
BD+49 2126	&	J12150885+4843574	&	12 15 08.84	&	+48 43 57.25	&	K6	&	0.65	&	22.46	 $\pm$ 	0.04	&	6469	 $^{+	4817	 }_{-	4366	}$ &	 ... 	&	...	&	 ... 	\\
BD+05 2767	&	J13342150+0440026	&	13 34 21.50	&	+04 40 02.63	&	K5	&	0.70	&	20.44	 $\pm$ 	0.03	&	6123	 $^{+	4735	 }_{-	4189	}$ &	 ... 	&	...	&	 ... 	\\
HD 128311	&	J14360055+0944474	&	14 36 00.56	&	+09 44 47.45	&	K3	&	0.80	&	16.34	 $\pm$ 	0.02	&	2672	 $^{+3962	 }_{-	1929	}$ &	11.54	 $\pm$ 	1.15	&	7	&	0.702	 	\\
BD+47 2936	&	J19505021+4804508	&	19 50 50.24	&	+48 04 51.09	&	K4	&	0.74	&	37.81	 $\pm$ 	0.03	&	1639	 $^{+	3891	 }_{-	76	}$ &	19.08	 $\pm$ 	0.06	&	8	&	0.778	 	\\
WASP-80	&	J20124017-0208391	&	20 12 40.16	&	-02 08 39.19	&	K7	&	0.61	&	49.86	 $\pm$ 	0.12	&	6067	 $^{+	4328	 }_{-	3238	}$ &	 ... 	&	...	&	 ... 	\\
WASP-69	&	J21000618-0505398	&	21 00 06.19	&	-05 05 40.03	&	K5	&	0.70	&	50.03	 $\pm$ 	0.13	&	5442	 $^{+	4919	 }_{-	3679	}$ &	 ... 	&	...	&	 ... 	\\
HD 201091	&	J21065341+3844529	&	21 06 53.93	&	+38 44 57.89	&	K5	&	0.70	&	3.50	 $\pm$ 	0.00	&	4803	 $^{+	4534	 }_{-	1238	}$ &	35.37	 $\pm$ 	3.54	&	9	&	1.889		\\
WASP-29	&	J23513108-3954241	&	23 51 31.08	&	-39 54 24.25	&	K4	&	0.74	&	87.82	 $\pm$ 	0.31	&	4003	 $^{+	5016	 }_{-	2876	}$ &	 ... 	&	...	&	 ... 	\\
\enddata
\tablerefs{		
(1)		\citet{Messina2010};
(2)		\citet{Jayasinghe2019};
(3)		\citet{Douglas2019};
(4)		\citet{Kiraga2013};
(5)		\citet{Astudillo-Defru2017};
(6)		\citet{Lu2019};
(7)		\citet{Strassmeier2000};
(8)		\citet{McQuillan2014};
(9) \citet{Mamajek2008}.}		
\tablenotetext{a}{Distance measurements from \textit{Gaia} EDR3 \citep{gaiadr3}.}
\tablenotetext{b}{Ages either from the literature for moving group members (\S\ref{sec:target_selection}) or derived from the \texttt{stardate} program by \citet{Angus2019} for field stars. See \S\ref{sec:ages, prots, rossby} for discussion. }
\tablenotetext{c}{Errors reported where specified by the original authors, otherwise assumed to be 10\% of the rotation period.}
\tablenotetext{d}{Identified in this work as a spectroscopic binary and was therefore not included in the analysis. See \S\ref{sec:data reduction} for discussion.}
\end{deluxetable}
\end{longrotatetable}
% If "omit" error, probably from vspace in the wrong place. Can't submit paper until that's fixed and no red errors. 
\startlongtable
\begin{deluxetable}{l c c c c c}
\centering
\tabletypesize{\tiny}
\tablecaption{\normalsize{Summary of observations.} \label{tab:obs_info}}
\tablehead{
\colhead{Star} \vspace{-0.3cm} & \colhead{Grating} & \colhead{$T_{exp}$} & \colhead{First} & \colhead{Last} & \colhead{Program}\\
\colhead{} \vspace{-0.3cm} & \colhead{} & \colhead{} & \colhead{Observation} & \colhead{Observation} & \colhead{HST \#}\\
\colhead{} \vspace{-0.3cm} & \colhead{} & \colhead{[s]} & \colhead{[MM/DD/YY]} & \colhead{[MM/DD/YY]} & \colhead{}\\
\vspace{-0.2cm}
}
\startdata
\multicolumn{6}{l}{Tucana-Horologium Members (40 Myr) } \\											
\hline											
HIP 1910	&	G130M	&	1636	&	08/30/17	&	…	&	14784	\\
	&	G160M	&	1502	&	08/30/17	&	…	&	14784	\\
	&	G230L	&	284	&	08/30/17	&	…	&	14784	\\
CT Tuc	&	G130M	&	5369	&	03/06/20	&	…	&	15955	\\
	&	G160M	&	1786	&	03/06/20	&	…	&	15955	\\
	&	G230L	&	562	&	03/06/20	&	…	&	15955	\\
CD-46	&	G130M	&	5173	&	04/27/20	&	…	&	15955	\\
	&	G160M	&	1837	&	04/27/20	&	…	&	15955	\\
	&	G230L	&	364	&	04/27/20	&	…	&	15955	\\
CD-35	&	G130M	&	4483	&	02/27/20	&	…	&	15955	\\
	&	G160M	&	3592	&	02/27/20	&	…	&	15955	\\
	&	G230L	&	578	&	02/27/20	&	…	&	15955	\\
CD-44 1173	&	G130M	&	10117	&	07/20/17	&	…	&	14784	\\
	&	G160M	&	10810	&	07/20/17	&	…	&	14784	\\
	&	G230L	&	4561	&	07/19/17	&	…	&	14784	\\
TYC 8083	&	G130M	&	4280	&	04/18/20	&	…	&	15955	\\
	&	G160M	&	2068	&	04/18/20	&	…	&	15955	\\
	&	G230L	&	644	&	04/18/20	&	…	&	15955	\\
TYC 8098	&	G130M	&	5932	&	06/18/20	&	…	&	15955	\\
	&	G160M	&	3512	&	06/18/20	&	…	&	15955	\\
	&	G230L	&	662	&	06/18/20	&	…	&	15955	\\
HIP 107345	&	G130M	&	5201	&	03/13/20	&	…	&	15955	\\
	&	G160M	&	1845	&	03/13/20	&	…	&	15955	\\
	&	G230L	&	662	&	03/13/20	&	…	&	15955	\\
J23261	&	G130M	&	1352	&	08/18/17	&	…	&	14784	\\
	&	G160M	&	1402	&	08/18/17	&	…	&	14784	\\
	&	G230L	&	606	&	08/18/17	&	…	&	14784	\\
UCAC3 13	&	G130M	&	6334	&	02/16/21	&	…	&	15955	\\
	&	G160M	&	1531	&	02/16/21	&	…	&	15955	\\
	&	G230L	&	1506	&	02/16/21	&	…	&	15955	\\
\hline											
\multicolumn{6}{l}{Hyades Members (650 Myr)} \\											
\hline											
J03510	&	G230L	&	88	&	09/29/19	&	…	&	15091	\\
J03524	&	G230L	&	868	&	08/07/19	&	…	&	15091	\\
J03550	&	G230L	&	208	&	09/18/19	&	…	&	15091	\\
J04070	&	G230L	&	395	&	07/29/19	&	…	&	15091	\\
J04081	&	G230L	&	777	&	09/22/19	&	…	&	15091	\\
J04082	&	G230L	&	271	&	08/24/19	&	…	&	15091	\\
J04172	&	G230L	&	678	&	09/06/19	&	…	&	15091	\\
J04240	&	G230L	&	934	&	11/01/19	&	…	&	15091	\\
J04293	&	G230L	&	1167	&	09/13/19	&	…	&	15091	\\
V925 Tau	&	G130M	&	5580	&	02/26/20	&	…	&	15955	\\
	&	G160M	&	2472	&	02/26/20	&	…	&	15955	\\
	&	G230L	&	1160	&	02/26/20	&	…	&	15955	\\
J04333	&	G230L	&	632	&	09/24/19	&	…	&	15091	\\
J04334	&	G230L	&	501	&	09/13/19	&	…	&	15091	\\
STKM 1-495	&	G130M	&	5556	&	02/23/20	&	…	&	15955	\\
	&	G160M	&	2806	&	02/23/20	&	…	&	15955	\\
	&	G230L	&	440	&	02/23/20	&	…	&	15955	\\
UCAC4 518	&	G130M	&	5564	&	09/07/20	&	…	&	15955	\\
	&	G160M	&	2979	&	09/07/20	&	…	&	15955	\\
	&	G230L	&	832	&	09/07/20	&	…	&	15955	\\
J04470	&	G230L	&	356	&	12/02/19	&	…	&	15091	\\
J04472	&	G230L	&	571	&	09/29/19	&	…	&	15091	\\
J04514	&	G230L	&	1179	&	09/15/19	&	…	&	15091	\\
\hline											
\multicolumn{6}{l}{Field Stars ($\approx$5 Gyr)} \\											
\hline											
CD-23	&	G130M	&	5580	&	12/26/20	&	…	&	15955	\\
	&	G160M	&	5370	&	12/26/20	&	…	&	15955	\\
	&	G230L	&	1522	&	12/26/20	&	…	&	15955	\\
BD+16	&	G130M	&	5578	&	03/03/20	&	…	&	15955	\\
	&	G160M	&	3428	&	03/03/20	&	…	&	15955	\\
	&	G230L	&	654	&	03/03/20	&	…	&	15955	\\
HD 266611	&	G130M	&	5626	&	02/25/20	&	…	&	15955	\\
	&	G160M	&	3617	&	02/25/20	&	…	&	15955	\\
	&	G230L	&	674	&	02/25/20	&	…	&	15955	\\
HD 85512	&	G130M	&	4631	&	03/14/15	&	…	&	13650	\\
	&	G160M	&	400	&	03/14/15	&	…	&	13650	\\
	&	G230L	&	458	&	03/14/15	&	…	&	13650	\\
BD+57	&	G130M	&	6044	&	05/02/20	&	…	&	15955	\\
	&	G160M	&	3910	&	05/02/20	&	…	&	15955	\\
	&	G230L	&	846	&	05/02/20	&	…	&	15955	\\
HD 99492	&	G130M	&	1920	&	03/31/17	&	…	&	14633	\\
BD+49	&	G130M	&	5836	&	06/10/20	&	…	&	15955	\\
	&	G160M	&	5626	&	06/10/20	&	…	&	15955	\\
	&	G230L	&	1614	&	06/10/20	&	…	&	15955	\\
BD+05	&	G130M	&	11103	&	06/12/20	&	…	&	15955	\\
	&	G160M	&	2668	&	06/12/20	&	…	&	15955	\\
	&	G230L	&	1738	&	06/12/20	&	…	&	15955	\\
HD 128311	&	G130M	&	1920	&	06/10/17	&	…	&	14633	\\
BD+47	&	G130M	&	38375	&	12/16/16	&	05/21/17	&	14767	\\
WASP-80	&	G130M	&	43273	&	07/22/18	&	06/28/19	&	14767	\\
WASP-69	&	G130M	&	38226	&	10/12/17	&	10/31/17	&	14767	\\
HD 201091	&	G130M	&	1300	&	03/28/10	&	…	&	11687	\\
WASP-29	&	G130M	&	39020	&	05/19/18	&	11/04/18	&	14767	\\
\enddata
\end{deluxetable}

In this paper, we present an examination of the evolution of upper atmosphere UV emission lines of 39 K stars, building upon the photometric K star evolution work of \citet{richey-yowell2019}, as well as expanding the M star UV spectral analysis of \citet{Loyd2021} into the K star regime. Using the Cosmic Origins Spectrograph (COS) on the \textit{Hubble Space Telescope} (\textit{HST}), we obtained spectra of K stars at ages 40 Myr \citep[][]{KrausTucHor}, 650 Myr \citep[][]{Martin2018}, and $\approx$5 Gyr (the average age of the field stars) in both the NUV (2000-3100 \AA) and FUV (1100-1800 \AA) in order to measure the UV temporal and rotation evolution of K stars.  In \S\ref{sec:observations}, we discuss the target selection strategy followed by a description of the methods for analysis in \S\ref{sec:analysis}. Finally, we present the results in \S\ref{sec:results} and discussion the implications for these conclusions in \S\ref{sec:discussion}. 

\section{Observations}\label{sec:observations}

\subsection{Target Selection}\label{sec:target_selection}

To explore the UV evolution of K stars, we selected stars at three distinct ages: young, intermediate, and old. The young star sample of 10 K stars was taken from the Tucana-Horologium young moving group (40 $\pm$ 5 Myr, \citealt{KrausTucHor}) and confirmed to be bona-fide members through their 3D space velocity motions and lithium depletion by \citet{KrausTucHor}. Similarly, our intermediate stellar sample of 15 K stars was selected from members of the Hyades moving group (650 $\pm$ 70 Myr, \citealt{Martin2018}), confirmed through 3D space velocity kinematics by \citet{goldmanhyades}. Finally, we identified 14 field stars within 30 pc using the SIMBAD database and assumed an age of 5 Gyr for our field age sample, as this is the average age of the stellar field. 

K stars with the same spectral type will have different masses at different ages. Therefore, we must consider the evolution of stars of similar mass in time. We estimated the masses of the stars using Table 1 of \citet{richey-yowell2019}, which were calculated from the spectral type to effective temperature transformations of \citet{Pecaut2013} and the model isochrones of \citet{Baraffe2015}. We select stars only from 0.6 -- 0.8 M$_{\odot}$ to be included in our samples. A summary of the targets is presented in Table \ref{tab:target_info}. 

\subsection{Hubble Space Telescope Observations}\label{sec:hst_observations}

The 39 targets were observed in the NUV and/or FUV with the Hubble Space Telescope (HST) Cosmic Origins Spectrograph (COS) instrument. Of these, 16 are new observations from GO-15955 (PI: Richey-Yowell) and 14 are new observations from SNAP-15091 (PI: Agüeros); the remaining are publicly archived observations from GO-14767 (PI: Sing), GO-14784 (PI: Shkolnik), GO-13650 (PI: France), SNAP-14633 (PI: France), and SNAP-11687 (PI: Ayres). The \textit{HST} program number is identified with each target in Table \ref{tab:obs_info}. 

For the new targets from GO-15955 presented in this paper, each was visited once by \textit{HST} and was exposed in the COS G230L (NUV), G160M (FUV), and G130M (FUV) gratings, with central wavelengths 2950 \AA, 1577 \AA, and 1291 \AA \, respectively. Total exposure times ranged from $\approx$250 s to $\approx$12,000 s depending on the brightness of the object. Time between exposures were typically $\approx$100 s or $\approx$45 minutes if the Earth was occulting the target during HST's orbit. The individual exposures were dithered on the detector to reduce noise that may be introduced into the spectra from observations at a single location. For the new targets from SNAP-15091, each was visited once by \textit{HST} and exposed in the COS G230L grating with central wavelength 2950 \AA. Exposure times ranged from $\approx$90 s to $\approx$1,200 s. The additional archived targets were observed in at least one of the COS G230L, G160M, or G130M gratings, although the central wavelengths varied from program to program. A summary of the observations can be found in Table \ref{tab:obs_info}.

\section{Analysis}\label{sec:analysis} 

% If "omit" error, probably from vspace in the wrong place. Can't submit paper until that's fixed and no red errors. 

\begin{deluxetable}{l c c c }[t]
\centering
\tabletypesize{\scriptsize}
\tablecaption{\normalsize{Properties of the UV emission lines studied in this work.} \label{tab:line_info}}
\tablehead{
\colhead{Ion} \vspace{-0.3cm} & \colhead{Integration Region [\AA]} & \colhead{$\lambda_{rest}$ [\AA]} & \colhead{log$_{10}(T_{peak}$/[K])}\\
\vspace{-0.2cm}
}
\startdata
C III\tablenotemark{$\ast$}	&	1174.5 -- 1177.0	&	1174.93	&	4.7	\\
	&		&	1175.26	&	4.7	\\
	&		&	1175.59	&	4.7	\\
	&		&	1175.71	&	4.7	\\
	&		&	1175.99	&	4.7	\\
	&		&	1176.37	&	4.7	\\
Si III\tablenotemark{$\ast$}	&	1205.8 -- 1207.2	&	1206.51	&	4.7	\\
N V\tablenotemark{$\ast$}	&	1238.0 -- 1244.1	&	1238.82	&	5.2	\\
	&		&	1242.80	&	5.2	\\
C II\tablenotemark{$\ast$}	&	1333.3 -- 1336.8	&	1334.52	&	4.5	\\
	&		&	1335.71	&	4.5	\\
Si IV	&	1393.0 -- 1395.0	&	1393.76	&	4.9	\\
	&	1401.9 -- 1403.5	&	1402.77	&	4.9	\\
C IV	&	1547.0 -- 1552.0	&	1548.20	&	4.8	\\
	&		&	1550.77	&	4.8	\\
He II	&	1639.0 -- 1642.0	&	1640.4	&	4.9	\\
Mg II\tablenotemark{$\ast$}	&	2792.0 -- 2807.0	&	2796.35	&	3.6 -- 4.0	\\
	&		&	2808.53	&	3.6 -- 4.0	\\
\hline 
\multicolumn{2}{l}{ } & \multicolumn{2}{c}{Integration Regions [\AA]}\\
\multicolumn{2}{l}{FUV Pseudo-Continuum} & \multicolumn{2}{c}{1160-1172, 1250-1290, 1320-1330,}\\
\multicolumn{2}{l}{} & \multicolumn{2}{c}{1340-1390, 1406-1540, 1590-1635,}\\
\multicolumn{2}{l}{} & \multicolumn{2}{c}{1680-1740}\\
\multicolumn{2}{l}{NUV Pseudo-Continuum} & \multicolumn{2}{c}{2810-3200}\\
\enddata
\tablenotetext{\ast}{Outside of the GALEX NUV or FUV bandpasses.}
\end{deluxetable}

\begin{figure*}[t]
    \centering
        \includegraphics[width=\linewidth]{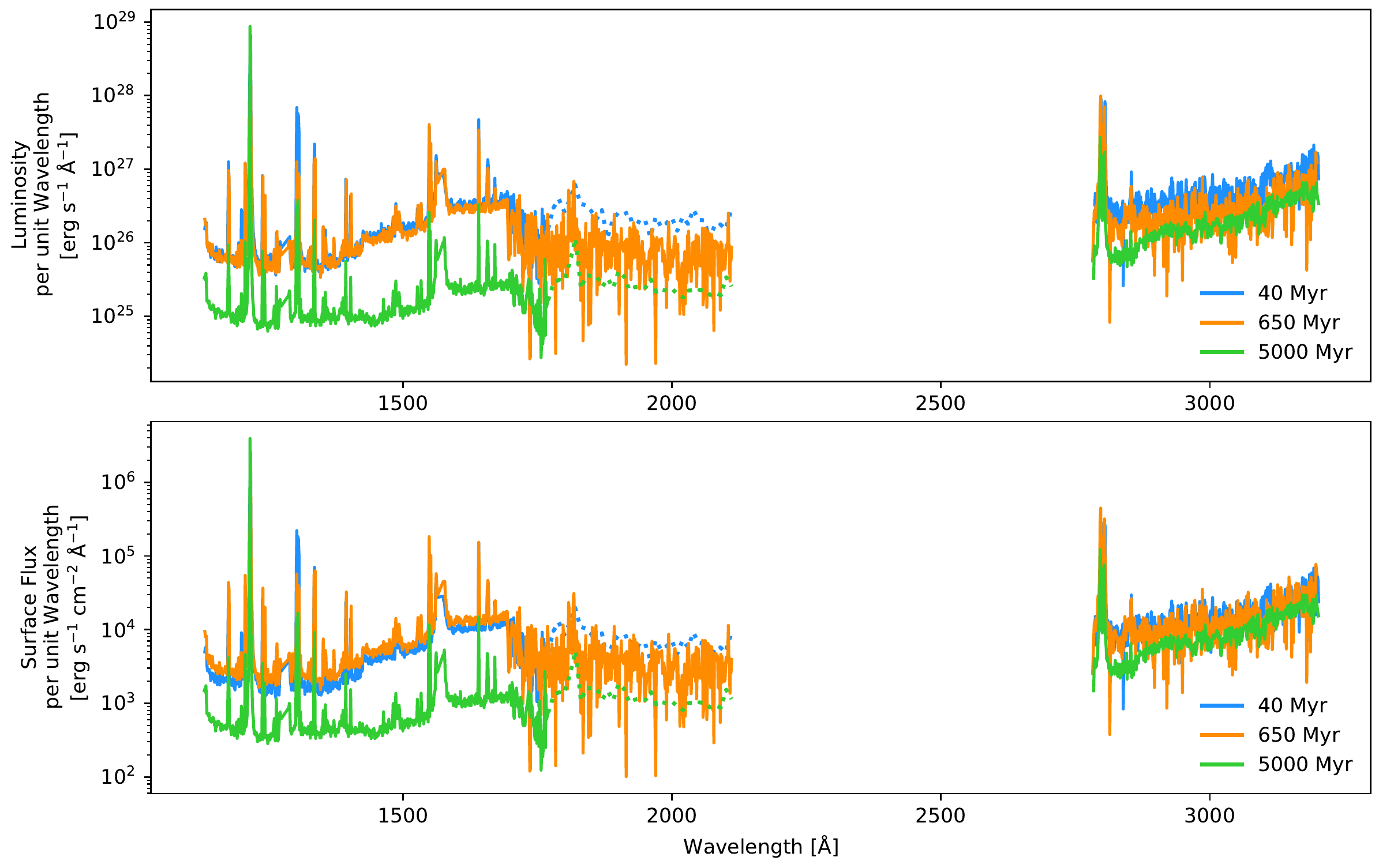}
    \caption{\textit{HST}/COS spectra of HIP 107345 (40 Myr), V925 Tau (650 Myr), and BD+49 2126 ($\approx$5 Gyr), as representative spectra of the entire sample. The upper panel shows the luminosity per unit wavelength of the spectra, while the lower panel shows the surface flux per unit wavelength. The data have been binned into 1-\AA \,intervals, except for the dotted portion of the spectra that represent 2$\sigma$ limits of low S/N data (S/N < 5 when binned to 10-\AA \,intervals).}  The 40 Myr and 650 Myr spectra appear quite similar to each other, while the field star spectrum is almost an order of magnitude lower in the FUV. Note that the Ly$\alpha$ (1216 \AA) and O I (1302 \AA) lines between the spectra are not comparable due to different levels of geocoronal contamination and ISM absorption. \label{fig:spectra}
\end{figure*}

\begin{figure*}[t]
    \centering
        \includegraphics[width=\linewidth]{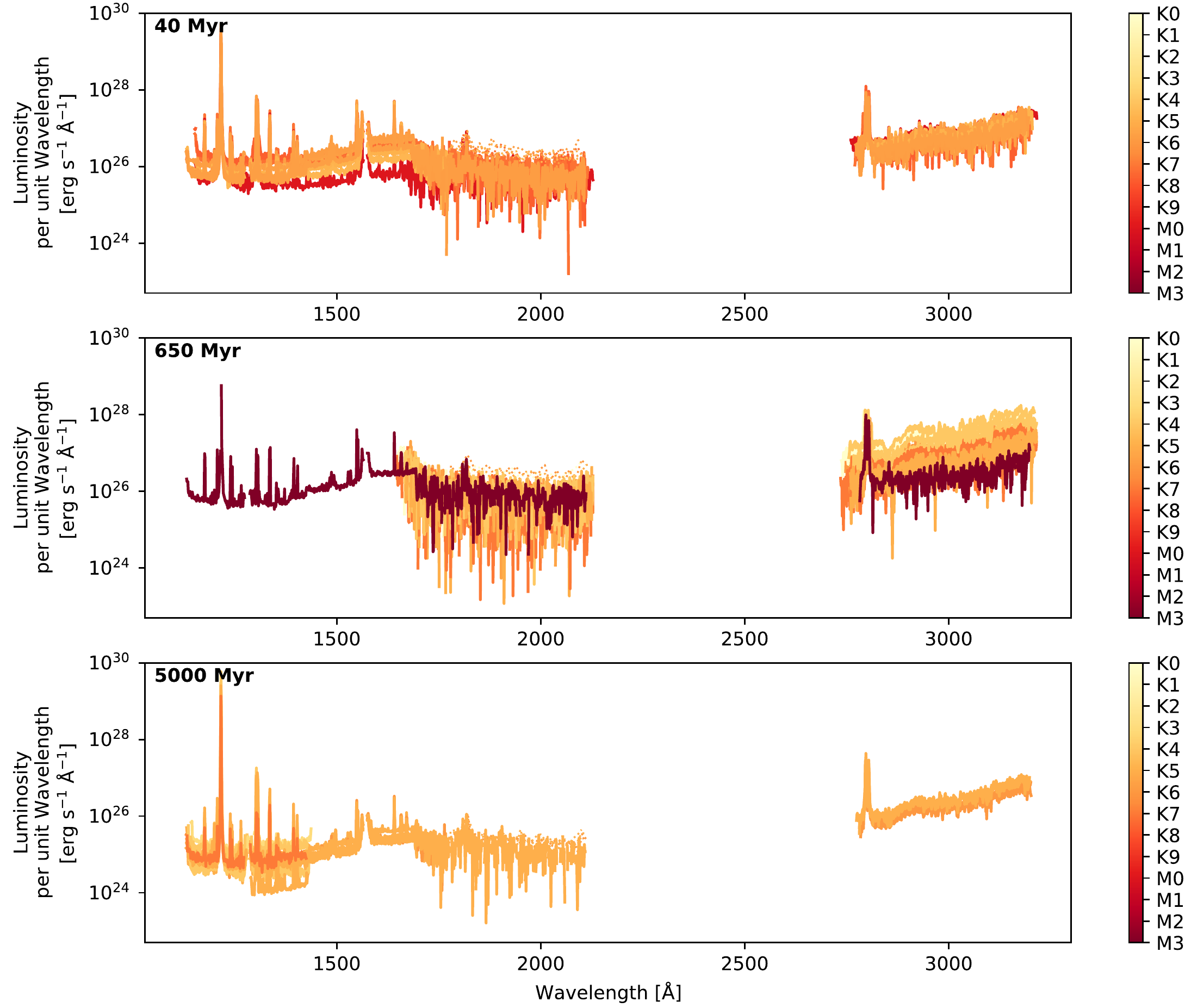}
    \caption{All \textit{HST} spectra, separated by age. Flares have been removed for the direct comparison of quiescent emission and the data have been binned into 1-\AA \,intervals, except for the dotted portion of the spectra that represent 2$\sigma$ limits of low S/N data (S/N < 5 when binned to 10-\AA \, intervals). Only one Hyades (650 Myr) member has data in the FUV. } There is clear residual variability between the spectra, particularly at the younger ages, partly due to a larger spread in spectral mass. Note that Ly$\alpha$ (1216 \AA) and O I (1302 \AA) lines between the spectra are not comparable due to different levels of geocoronal contamination and ISM absorption. \label{fig:comp_spectra}
\end{figure*}

\begin{figure}[t]
    \centering
        \includegraphics[width=\linewidth]{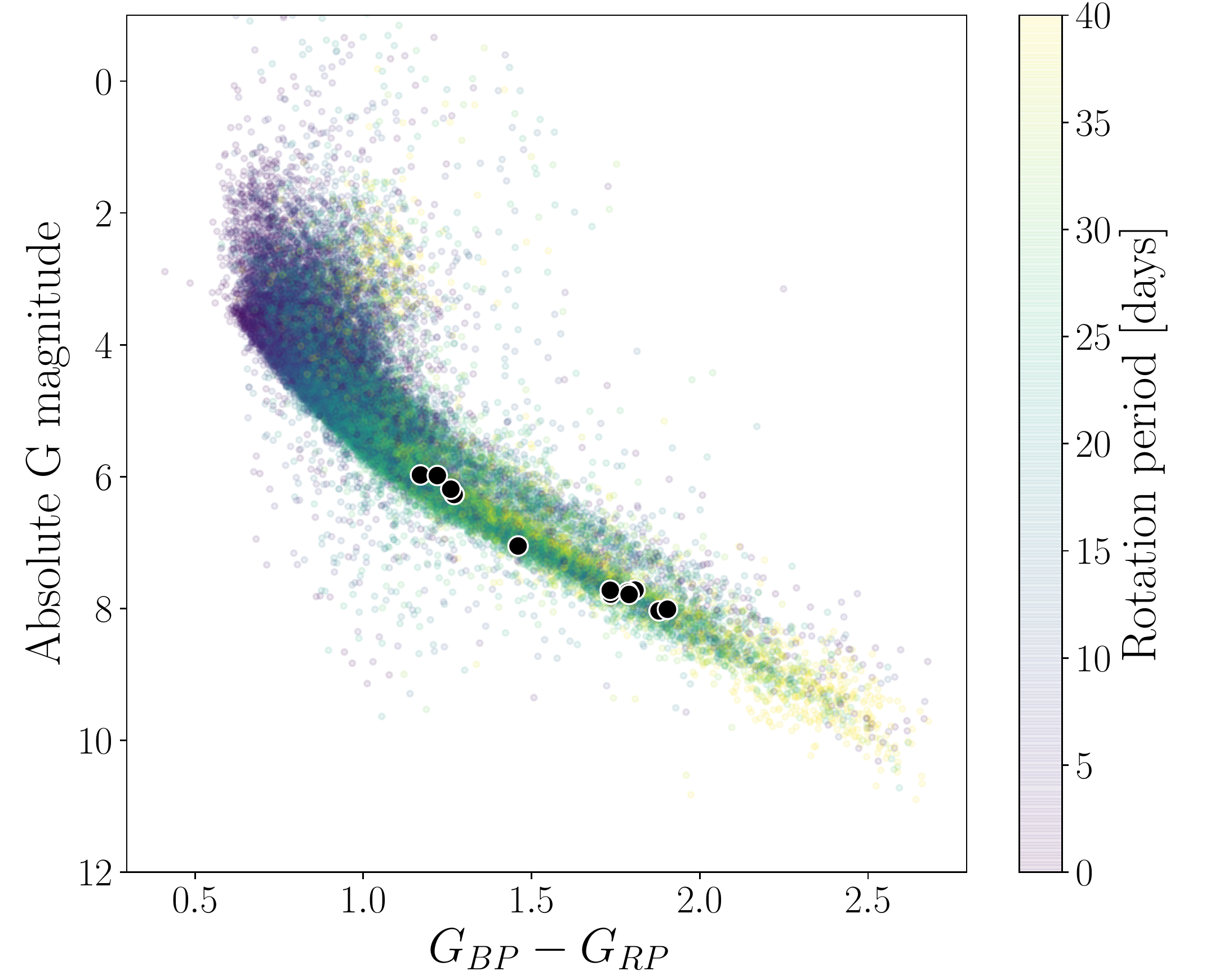}
    \caption{Color-magnitude diagram of \textit{Gaia} objects taken from the \texttt{stardate} package by \citet{Angus2019}. The field stars in this study are overlaid as black points. These stars were aged using \texttt{stardate} with the \textit{G}, \textit{BP}, and \textit{RP} magnitudes and parallaxes from \textit{Gaia} EDR3; the \textit{J}, \textit{H}, and \textit{K} magnitudes from 2MASS; and the rotation period where available.} \label{fig:hr}
\end{figure}

\subsection{Data Reduction}\label{sec:data reduction}

Initial data reduction was carried out through the \texttt{CALCOS} pipeline version 3.3.10. Since we are interested in the evolution of the quiescent UV flux of these stars, we removed any flares in the data. If a flare was found, we removed the flare by discarding the time region where the flare occurred using the \texttt{python} program \texttt{costools}.
The remaining data were then re-run through the \texttt{CALCOS} pipeline to produce flux-calibrated 1D spectra. While these spectra ideally represents a quiescent stellar state, unresolved flares may still be present.

To identify flares, we created light curves by extracting the spectrum (in terms of [counts s$^{-1}$ \AA$^{-1}$]) from the time-tagged (\texttt{corrtag}) files and subtracting the background region (see Figure 3 of \citealt{Loyd2014}). After removing regions contaminated by Earth's geocoronal emission (such as Ly$\alpha$ and O I), we then applied the flux calibration from the original \texttt{x1d} spectrum to the time-tagged spectrum, to create a conversion from [counts s$^{-1}$ \AA$^{-1}$] to [erg s$^{-1}$ cm$^{-2}$ \AA$^{-1}$]. After this, we integrated over the bandpass to create a flux-calibrated light curve for each target binned in 5 second intervals. A dozen flares above 3$\sigma$ were identified and will be analyzed separately in a future paper. No additional flare candidates were identified by eye.

For spectra that were observed during multiple visits, the data were coadded. We used the \texttt{coaddx1d}\footnote{\url{https://github.com/cosmonaut/python-coaddx1d}} \citep[][]{Danforth2010, Danforth2016} code originally in IDL and translated into python by Nicholas Nell to coadd the data, weighting by error. 

The final spectra were shifted to the rest frame of the star and reviewed individually. Examples representative of the spectra at each age are shown in Figure \ref{fig:spectra}. 

Two of the observed targets from GO-15955, STKM 1-495 and UCAC4 518-008871 showed clear binarity in our spectroscopic observations. Additionally, archival Keck NIRC2 images (Proposal I.D. \#H246N2, PI: Gaidos) confirm that STKM 1-495 is in fact a binary system. These two targets were thus not considered in our study due to both blending and potential increased activity due to tidal and/or magnetic interactions between their companions.

\subsection{Measuring Line and Continua Fluxes}

We analyze the strongest chromospheric and transition region emission features in the COS bandpasses: C III, Si III, N V, C II, Si IV, C IV, He II, and Mg II. These lines were also analyzed in \citet{Loyd2021} and thus we can make direct comparisons with the UV evolution of early M stars. We list the rest wavelength and formation temperature in Table \ref{tab:line_info}, as well as the wavelength range over which we integrate to calculate the surface fluxes.

For more accurate line fluxes, we used the \texttt{astropy} module \texttt{specutils} to fit the continuum immediately outside the spectral regions identified in Table \ref{tab:line_info} with a cubic Chebyshev polynomial and subtract the fit. The accuracy of these fits were confirmed by eye for each line of each target. 

We additionally analyze the NUV continuum as well as the FUV continuum of each spectrum. The continua in these regions are dominated by multiple weak absorption and emission lines and therefore are more accurately considered to be "psuedo-continua". To calculate the pseudo-continua, we consider the same regions as \citet{Loyd2021}, which cumulatively cover 251 \AA. The wavelengths of the regions over which we integrated are additionally found in Table \ref{tab:line_info}. These regions exclude areas of strong emission lines not considered in this study and non-stellar background contamination such as geo-coronal emission features. For the archived field-age targets which only had data in one of the two FUV gratings, we take the ratio of the target pseudo-continuum available to the same region in HD 266611 (the target with the FUV pseudo-continuum flux closest to the average of the FUV pseudo-continuum flux for the entire sample), and multiply the FUV pseudo-continuum flux of HD 266611 by that ratio. This assumes that the shape of the continuum does not vary significantly between targets, as is the case in Figure \ref{fig:comp_spectra}.

In order to calculate the surface flux from the observed flux, we estimated the radius of the star first using the spectral type to effective temperature relationship in \citet{Pecaut2013} and then using the effective temperature to mass and radius relationships in \citet{Baraffe2015}.

\begin{figure*}[t]
    \centering
        \includegraphics[width=\linewidth, ]{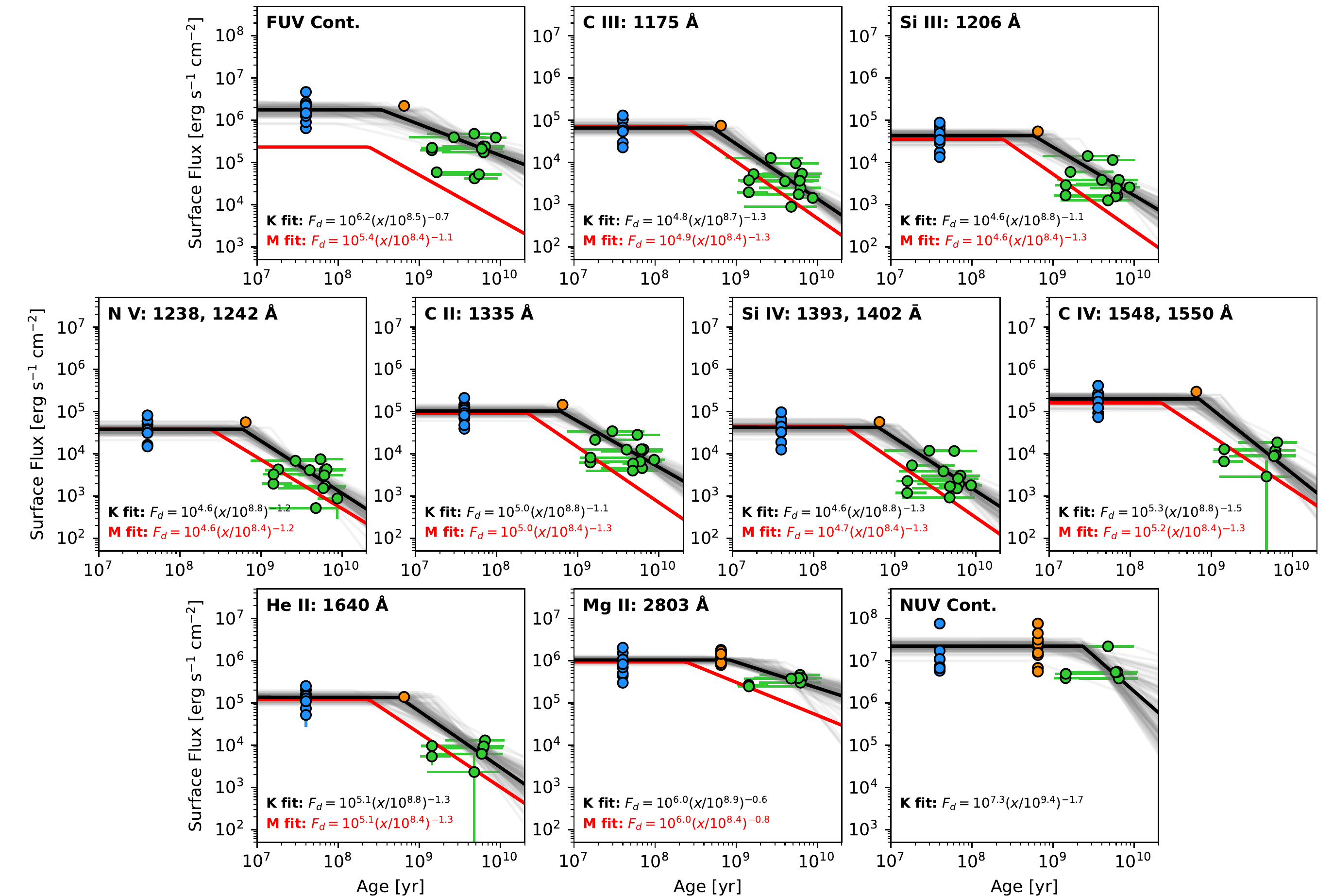}
    \caption{Surface flux as a function of stellar age for the 39 K stars in our sample. 40 Myr stars are in blue, 650 Myr in orange, and field stars in green. Only one Hyades (650 Myr) member has data in the FUV. The ages of the field stars were determined using \texttt{stardate} \citep[][]{Angus2019}. Model fits to Equation \ref{eq: 1} are shown in gray. Single gray lines well away from the average are all outliers compared to the total data set. The averages of the fits are shown in black and the resulting equations for the power law decline are shown, presented in full in Table \ref{tab:line_fits}. The red lines are the model fits for early M stars from \citet{Loyd2021}.}\label{fig:age}
\end{figure*}

\begin{figure*}[t]
    \centering
        \includegraphics[width=\linewidth]{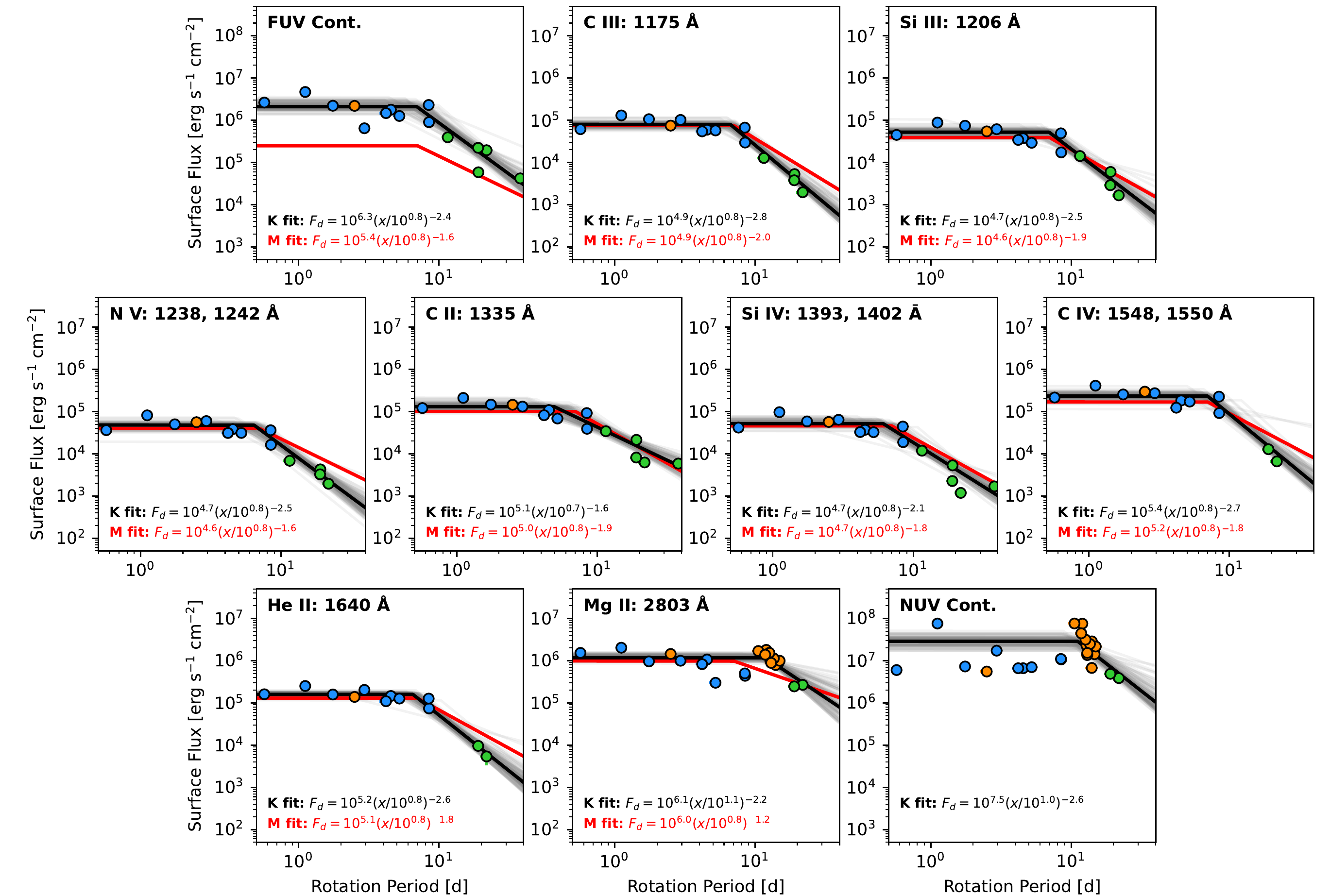}
    \caption{Same as Figure \ref{fig:age}, but for the evolution of the surface flux with rotation period.} \label{fig:prot}
\end{figure*}

\begin{figure*}[t]
    \centering
        \includegraphics[width=\linewidth]{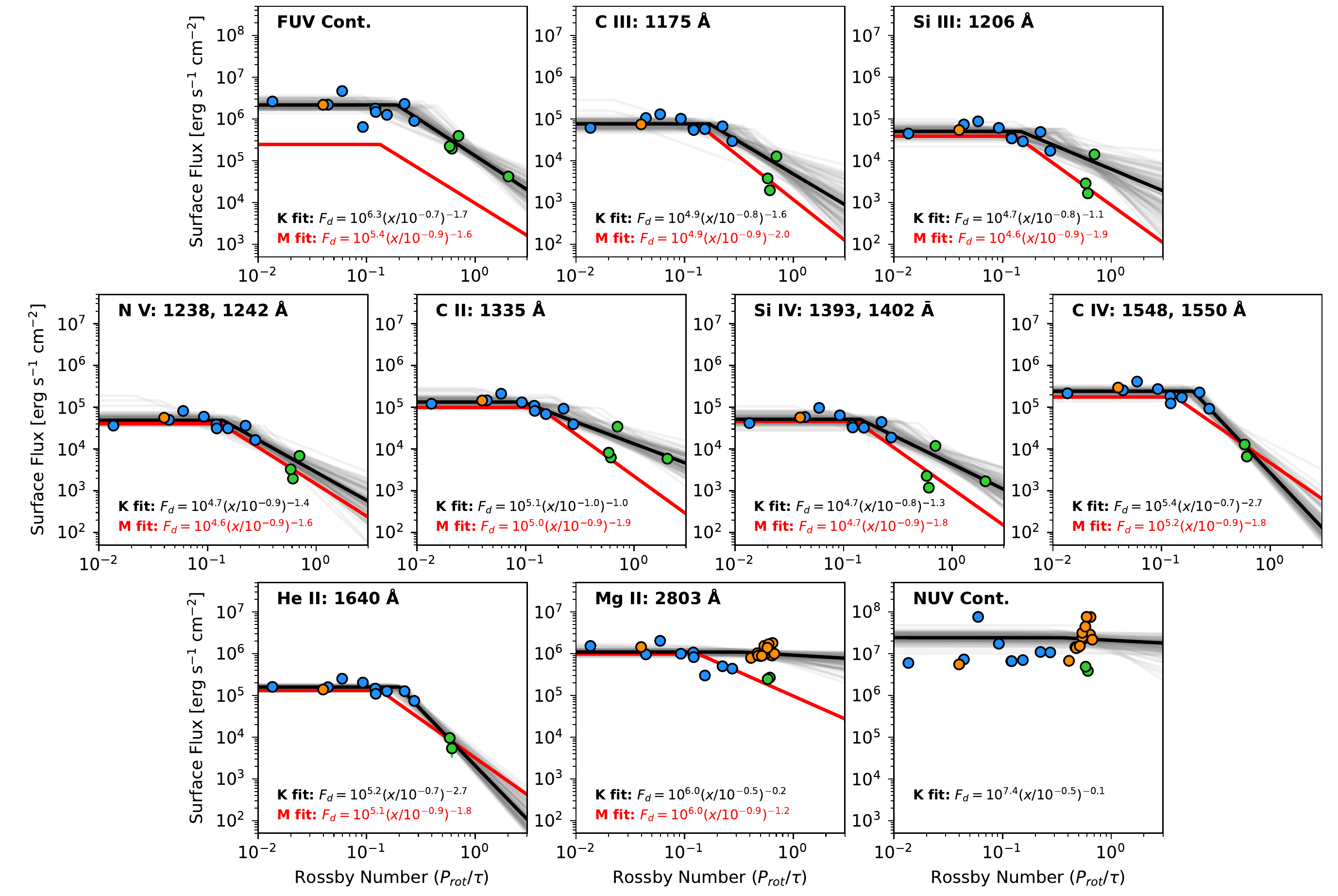}
    \caption{Same as Figure \ref{fig:age}, but for the evolution of the surface flux with Rossby number.} \label{fig:rossby}
\end{figure*}

\subsection{Determining Ages, Rotation Periods, and Rossby Numbers}\label{sec:ages, prots, rossby}

While the ages of the moving group members are well constrained, the ages of field stars are typically estimated from isochrones and gyrochrones. We utilized the \texttt{stardate} package by \citet{Angus2019} to determine these ages, providing the $G, BP,$ and $RP$ magnitudes and parallaxes from $Gaia$ EDR3 \citep[][]{gaiadr3}, the $J, H, $ and $K$ magnitudes from 2MASS \citep[][]{2mass}, and the rotation period if available. A summary of these data are seen in Figure \ref{fig:hr}.

The rotation periods were gathered from the literature where available. For stars with no reported error in their rotation period, we assume 10\%. For those without measured rotation periods, we carried out a Lomb-Scargle \citep[][]{lomb, scargle} search of $TESS$ lightcurves using the program \texttt{lightkurve} \citep[][]{Lightkurve}. The \texttt{lightkurve} periodograms test frequencies up to the Nyquist frequency, using grid steps of $1 / \textrm{observation time}$. UCAC3 13-31217, HD 266611, HD 85512, BD+57 1274, BD+49 2126, and BD+05 2767 had \textit{TESS} observations; however, no clear variability was identified, which is not uncommon for older stars.

The Rossby number is defined as the ratio of a star's rotation period to its convective turnover time (i.e. the time it takes for circulation within a convective cell): $R_o = P_{rot}/\tau$. This number was first shown by \citet{Noyes1984} to be correlated to the magnetic activity of stars, and therefore should relate directly to the chromospheric, transition region, and coronal surface fluxes of stars which are driven by this magnetic activity. To calculate the convective turnover time, we use the empirical relationship between $V-K_s$ color and the turnover timescale of convection by \citet{Wright2018}. For two stars in our data without $V-K_s$ colors, we instead use the mass -- convective turnover time relationship, again established by \citet{Wright2018}. These were found to produce consistent Rossby numbers with the rest of the data set.

The complete list of estimated stellar ages, measured rotation periods, and derived Rossby numbers can be found in Table \ref{tab:target_info}.

\subsection{Model Fitting}
To fit the surface flux, $F$, as a function of age, rotation period, or Rossby number, we adopt a piecewise model that includes a period of saturated (constant) activity, followed by a power law decline, as utilized in other chromospheric studies \citep[e.g.][]{Wright2011, jackson12J, Loyd2021, Pineda2021}:

\begin{equation}\label{eq: 1}
    F = 
    \begin{cases} 
      F_{sat}, & x \leq x_{sat} \\
      F_{sat} (x/x_{sat})^{-\alpha}, & x > x_{sat} \\
    \end{cases}
\end{equation}

\noindent where the data $x$ can be age, rotation period, or Rossby number; and the free parameters are the surface flux at saturation, $F_{sat}$; end of saturation, $x_{sat}$; and the power-law index, $\alpha$. We additionally include a hyperparameter, $f$, representing a fractional amount by which the error is underestimated due to intrinsic variation among stars. We fit this model using the Markov Chain Monte Carlo (MCMC) package \texttt{emcee} \citep{Foreman-Mackey2013}. We use uniform priors on the the power-law index, restricting it to $ 0 \leq \alpha \leq 3$, as values between this range are what is known for M stars \citep[][]{Loyd2021} and G stars \citep[][and references therein]{Gudel2007}. Placing a prior on the power-law index additionally helps constrain the age at which saturation would occur. We ran 10,000 steps, taking the result to be the 50th percentile value, and the errors as the 16th and 84th percentile values.
These results can be found in Table \ref{tab:line_fits}.

\section{Results}\label{sec:results}

Figure \ref{fig:spectra} shows examples of K star spectra representative of each sample at the three different ages used in this study: 40 Myr, 650 Myr, and $\approx$5 Gyr. The 40 Myr and 650 Myr spectra appear to be quite similar in both luminosity and surface flux per unit wavelength, while the 5 Gyr spectrum is an order of magnitude lower in the FUV, and half an order of magnitude lower in the NUV. 

These results are notably different than those of the early M stars presented in \citet{Loyd2021}. In their work, the authors show that the early M dwarf FUV spectra at each age are an order of magnitude separate from each other, with the exception of the rotation-period-normalized spectra, which lie on top of each other. In the NUV, the luminosity  and surface flux per unit wavelength of the Hyades and field-age spectra are similar, while the young star spectrum is an order of magnitude larger. However, the FUV spectra of both the 40 Myr and 650 Myr K stars have similar luminosity densities and surface flux densities to the young M stars. 

As well as comparing the broadband characteristics of the spectra, we also analyze specific spectral lines that probe the chromosphere and transition region of the stars and see how those evolve in time. Figure \ref{fig:age} shows the evolution of the surface flux of the strong UV emission lines C III, Si III, N V, C II, Si IV, C IV, He II, Mg II, and the FUV and NUV pseudo-continua as a function of stellar age. While the ages are well constrained for the moving group samples, large errorbars persist for the ages of the field age stars, even for those with measured rotation periods to aid in gyrochronological and isocohronal dating, as discussed in Section \ref{sec:analysis}. We again see in each emission line region that the K dwarfs appear to have a prolonged saturation period as compared to M stars, which fall off by only a couple hundred million years old. While the evolution for M stars is now quite well known, the age at which K stars actually do begin to decline in UV activity remains poorly constrained. 

Most (32/39) of our targets have measured rotation periods. The rotation periods of the 40 Myr and 650 Myr samples overlap slightly, while the field-age stars are distinct in their longer rotation. Figure \ref{fig:prot} shows the surface flux evolution of the emission lines and pseudo-continuum as a function of rotation period for those with known rotation periods. The surface flux remains roughly constant from  $P_{rot} \approx 0.5 - 10$ days for all emission lines and the pseudo-continuum, at which point the flux decays. 

Since our stars are of similar masses, the evolution of the surface flux of the emission lines with Rossby number is similar to that of the rotation period evolution, as seen in Figure \ref{fig:rossby}. The Rossby number at which the flux starts to decline is $Ro \sim 0.2$. Since $Ro$ is the rotation period normalized by the convective turnover time, it is more appropriate to compare the K and M stars in terms of $Ro$ instead of rotation. Interestingly, we see that the emission line evolution fits between the partially convective K stars and early M stars are nearly identical to each other. This implies that the evolution of K dwarf rotation period with age is distinct from that of early M stars and is the primary driver of the observed differences in activity evolution with age. 

In each of these cases, the slope of the Mg II surface flux evolution is more shallow than for the other emission lines. Mg II is the only chromospheric line analyzed here, while the others map to the transition region of the star. The difference in the slopes may represent a difference in heating mechanism between the chromosphere and the transition region. However, ISM absorption may also play a role, since the Tuc-Hor and Hyades moving groups lie at a distance of 40 -- 60 pc and thus may have a greater Mg column density than for the field stars, which range from 3 -- 50 pc. A greater Mg absorption in the younger stars may artificially create a descreased slope between the young and field-aged stars. For direct comparison among all of the lines, the model fits for the UV surface flux evolution with age, rotation period, and Rossby number are presented in Table \ref{tab:line_fits}.

% If "omit" error, probably from vspace in the wrong place. Can't submit paper until that's fixed and no red errors. 																									
																									
\begin{deluxetable*}{l c c c c}[t]																									
\centering																									
\tabletypesize{\footnotesize}																									
\tablecaption{\normalsize{Results of the MCMC fits to the data seen in Figures \ref{fig:age}, \ref{fig:prot}, and \ref{fig:rossby}. } \label{tab:line_fits}}																									
\tablehead{																									
\colhead{Line} \vspace{-0.3cm} & \colhead{log$_{10}(x_{sat}$)} & \colhead{log$_{10}(F_{sat}$)} & \colhead{$\alpha$} & \colhead{log($f$)}\\																									
\colhead{} & \colhead{[year]} & \colhead{[erg s$^{-1}$ cm$^{-2}$]} & \colhead{}\\																									
\vspace{-0.5cm}																									
}																									
\startdata																									
\hline																									
Age \\																									
\hline																									
Mg II	& $	8.920	_{-	0.168	} ^{+	0.275	} $ & $	6.016	_{-	0.041	} ^{+	0.041	} $ & $	0.613	_{-	0.150	} ^{+	0.277	} $ & $	-0.859	_{-	0.153	} ^{+	0.174	} $ \\
C II	& $	8.779	_{-	0.130	} ^{+	0.126	} $ & $	5.012	_{-	0.070	} ^{+	0.064	} $ & $	1.090	_{-	0.161	} ^{+	0.195	} $ & $	-0.649	_{-	0.222	} ^{+	0.253	} $ \\
Si III	& $	8.760	_{-	0.153	} ^{+	0.148	} $ & $	4.637	_{-	0.077	} ^{+	0.071	} $ & $	1.142	_{-	0.185	} ^{+	0.230	} $ & $	-0.534	_{-	0.238	} ^{+	0.269	} $ \\
C III	& $	8.708	_{-	0.129	} ^{+	0.136	} $ & $	4.818	_{-	0.075	} ^{+	0.066	} $ & $	1.298	_{-	0.182	} ^{+	0.223	} $ & $	-0.592	_{-	0.237	} ^{+	0.266	} $ \\
Si IV	& $	8.797	_{-	0.149	} ^{+	0.148	} $ & $	4.626	_{-	0.082	} ^{+	0.072	} $ & $	1.250	_{-	0.210	} ^{+	0.277	} $ & $	-0.496	_{-	0.238	} ^{+	0.275	} $ \\
He II	& $	8.753	_{-	0.179	} ^{+	0.194	} $ & $	5.131	_{-	0.064	} ^{+	0.054	} $ & $	1.331	_{-	0.261	} ^{+	0.356	} $ & $	-0.885	_{-	0.297	} ^{+	0.327	} $ \\
C IV	& $	8.847	_{-	0.111	} ^{+	0.153	} $ & $	5.299	_{-	0.079	} ^{+	0.066	} $ & $	1.533	_{-	0.244	} ^{+	0.352	} $ & $	-0.635	_{-	0.266	} ^{+	0.316	} $ \\
N V	& $	8.778	_{-	0.124	} ^{+	0.123	} $ & $	4.580	_{-	0.076	} ^{+	0.069	} $ & $	1.239	_{-	0.182	} ^{+	0.220	} $ & $	-0.580	_{-	0.235	} ^{+	0.274	} $ \\
FUV Cont.	& $	8.534	_{-	0.258	} ^{+	0.215	} $ & $	6.246	_{-	0.087	} ^{+	0.082	} $ & $	0.736	_{-	0.138	} ^{+	0.153	} $ & $	-0.354	_{-	0.226	} ^{+	0.265	} $ \\
NUV Cont.	& $	9.368	_{-	0.309	} ^{+	0.112	} $ & $	7.341	_{-	0.089	} ^{+	0.080	} $ & $	1.688	_{-	0.901	} ^{+	0.875	} $ & $	-0.010	_{-	0.206	} ^{+	0.247	} $ \\
\hline																									
Rotation Period \\																									
\hline																									
Mg II	& $	1.073	_{-	0.034	} ^{+	0.029	} $ & $	6.070	_{-	0.045	} ^{+	0.046	} $ & $	2.217	_{-	0.683	} ^{+	0.506	} $ & $	-0.945	_{-	0.170	} ^{+	0.188	} $ \\
C II	& $	0.694	_{-	0.158	} ^{+	0.115	} $ & $	5.117	_{-	0.069	} ^{+	0.077	} $ & $	1.577	_{-	0.314	} ^{+	0.325	} $ & $	-0.938	_{-	0.224	} ^{+	0.259	} $ \\
Si III	& $	0.842	_{-	0.061	} ^{+	0.049	} $ & $	4.717	_{-	0.067	} ^{+	0.064	} $ & $	2.526	_{-	0.403	} ^{+	0.302	} $ & $	-0.848	_{-	0.229	} ^{+	0.281	} $ \\
C III	& $	0.823	_{-	0.039	} ^{+	0.034	} $ & $	4.901	_{-	0.060	} ^{+	0.055	} $ & $	2.773	_{-	0.276	} ^{+	0.161	} $ & $	-0.979	_{-	0.238	} ^{+	0.278	} $ \\
Si IV	& $	0.786	_{-	0.094	} ^{+	0.068	} $ & $	4.714	_{-	0.072	} ^{+	0.073	} $ & $	2.088	_{-	0.317	} ^{+	0.265	} $ & $	-0.767	_{-	0.239	} ^{+	0.277	} $ \\
He II	& $	0.814	_{-	0.048	} ^{+	0.039	} $ & $	5.200	_{-	0.039	} ^{+	0.038	} $ & $	2.642	_{-	0.351	} ^{+	0.245	} $ & $	-1.524	_{-	0.311	} ^{+	0.343	} $ \\
C IV	& $	0.844	_{-	0.047	} ^{+	0.046	} $ & $	5.369	_{-	0.069	} ^{+	0.062	} $ & $	2.730	_{-	0.407	} ^{+	0.197	} $ & $	-0.875	_{-	0.262	} ^{+	0.310	} $ \\
N V	& $	0.807	_{-	0.069	} ^{+	0.049	} $ & $	4.682	_{-	0.060	} ^{+	0.058	} $ & $	2.466	_{-	0.395	} ^{+	0.307	} $ & $	-0.995	_{-	0.243	} ^{+	0.277	} $ \\
FUV Cont.	& $	0.841	_{-	0.073	} ^{+	0.060	} $ & $	6.322	_{-	0.084	} ^{+	0.083	} $ & $	2.428	_{-	0.348	} ^{+	0.301	} $ & $	-0.598	_{-	0.232	} ^{+	0.283	} $ \\
NUV Cont.	& $	1.046	_{-	0.034	} ^{+	0.030	} $ & $	7.459	_{-	0.097	} ^{+	0.092	} $ & $	2.596	_{-	0.620	} ^{+	0.297	} $ & $	-0.160	_{-	0.211	} ^{+	0.266	} $ \\
\hline																									
Rossby Number \\																									
\hline																									
Mg II	& $	-0.506	_{-	0.502	} ^{+	0.164	} $ & $	6.040	_{-	0.048	} ^{+	0.054	} $ & $	0.152	_{-	0.105	} ^{+	0.252	} $ & $	-0.763	_{-	0.159	} ^{+	0.187	} $ \\
C II	& $	-1.025	_{-	0.192	} ^{+	0.262	} $ & $	5.123	_{-	0.090	} ^{+	0.103	} $ & $	0.976	_{-	0.184	} ^{+	0.288	} $ & $	-0.782	_{-	0.233	} ^{+	0.273	} $ \\
Si III	& $	-0.833	_{-	0.344	} ^{+	0.327	} $ & $	4.705	_{-	0.093	} ^{+	0.105	} $ & $	1.087	_{-	0.378	} ^{+	0.843	} $ & $	-0.618	_{-	0.244	} ^{+	0.306	} $ \\
C III	& $	-0.775	_{-	0.235	} ^{+	0.169	} $ & $	4.888	_{-	0.079	} ^{+	0.088	} $ & $	1.552	_{-	0.476	} ^{+	0.594	} $ & $	-0.727	_{-	0.252	} ^{+	0.294	} $ \\
Si IV	& $	-0.847	_{-	0.250	} ^{+	0.216	} $ & $	4.710	_{-	0.092	} ^{+	0.108	} $ & $	1.267	_{-	0.302	} ^{+	0.384	} $ & $	-0.620	_{-	0.239	} ^{+	0.291	} $ \\
He II	& $	-0.700	_{-	0.041	} ^{+	0.032	} $ & $	5.199	_{-	0.037	} ^{+	0.036	} $ & $	2.687	_{-	0.330	} ^{+	0.216	} $ & $	-1.563	_{-	0.306	} ^{+	0.344	} $ \\
C IV	& $	-0.710	_{-	0.040	} ^{+	0.038	} $ & $	5.381	_{-	0.057	} ^{+	0.055	} $ & $	2.746	_{-	0.336	} ^{+	0.182	} $ & $	-1.030	_{-	0.257	} ^{+	0.302	} $ \\
N V	& $	-0.860	_{-	0.198	} ^{+	0.167	} $ & $	4.689	_{-	0.077	} ^{+	0.088	} $ & $	1.449	_{-	0.358	} ^{+	0.498	} $ & $	-0.867	_{-	0.254	} ^{+	0.296	} $ \\
FUV Cont.	& $	-0.720	_{-	0.160	} ^{+	0.121	} $ & $	6.336	_{-	0.083	} ^{+	0.088	} $ & $	1.692	_{-	0.346	} ^{+	0.324	} $ & $	-0.663	_{-	0.248	} ^{+	0.289	} $ \\
NUV Cont.	& $	-0.474	_{-	0.382	} ^{+	0.136	} $ & $	7.385	_{-	0.103	} ^{+	0.091	} $ & $	0.138	_{-	0.103	} ^{+	0.247	} $ & $	0.015	_{-	0.221	} ^{+	0.280	} $ \\
\enddata																									
\end{deluxetable*}

\subsection{Comparison with \textit{GALEX} photometry}

\begin{figure*}[t]
    \centering
        \includegraphics[width=0.7\linewidth]{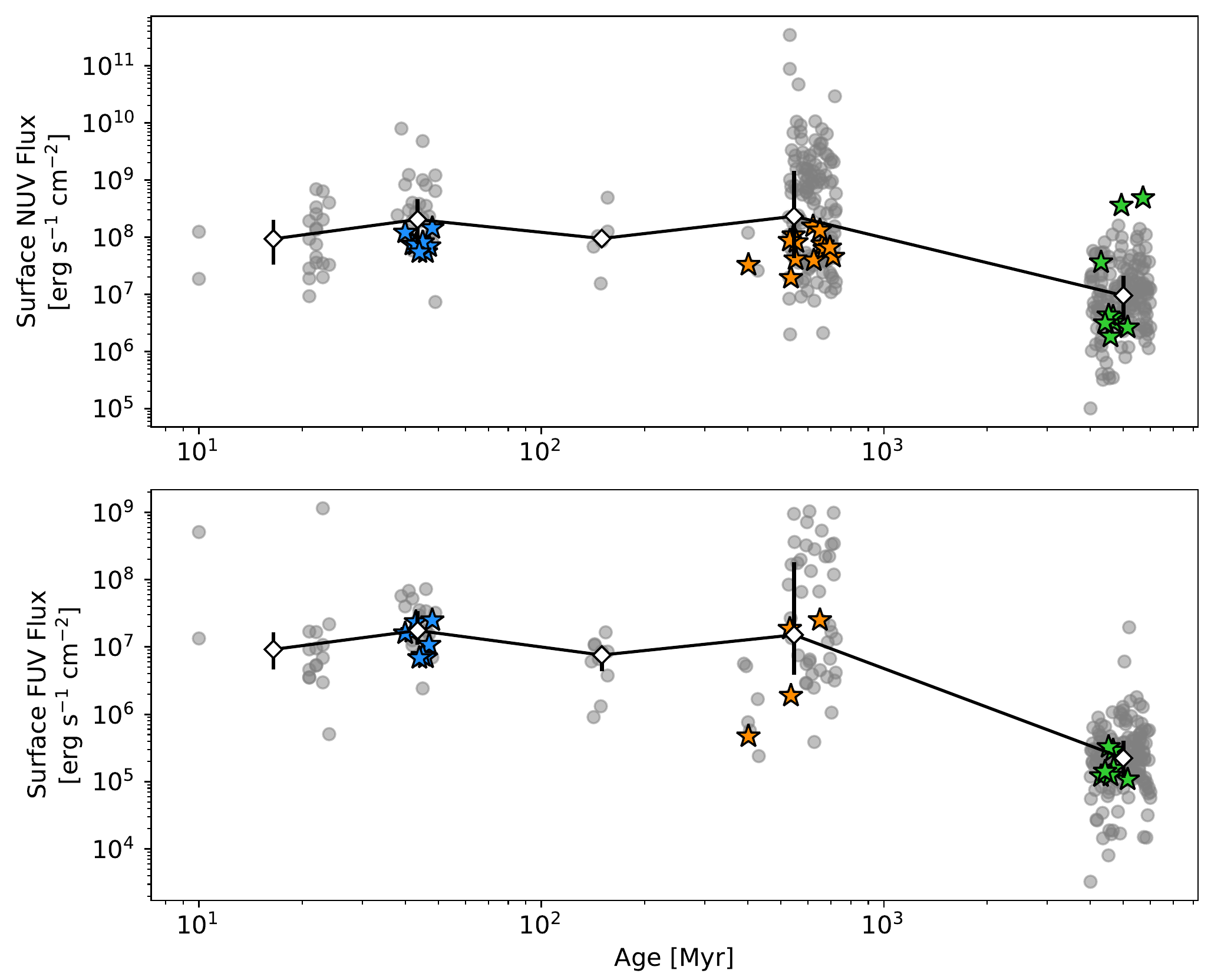}
    \caption{Relative GALEX photometric UV fluxes of the stars in our sample compared to the entire sample studied in \citet{richey-yowell2019}. Artificial age scatter has been added to both samples for clarity. The colored stars represent targets in our sample, whereas gray points are the entire photometric sample. White diamonds represent the median flux of that age bin, with the interquartiles (middle 50\%) presented as the error bars. Our targets typically lie within the interquartiles and do not appear to be unique to the total sample. } \label{fig:galex}
\end{figure*}

Using GALEX broadband UV photometry of 400 K stars ranging in age from 10 Myr to $\approx$5 Gyr, \citet{richey-yowell2019} saw a decrease (half an order of magnitude) in median flux by 650 Myr; however, the  interquartiles (the middle 50\% of the population) continued to overlap with the younger stars. Figure \ref{fig:galex} shows the relative GALEX UV surface fluxes of the stars in this work compared to their complete sample. The stars in this work are typical compared to other K stars and fall within the interquartiles of the GALEX sample. 

It should also be noted that the X-ray evolution of K stars \citep[e.g.][]{richey-yowell2019, jackson12J} starts to decline after 10 Myr, much earlier than seen in the UV, unlike for M stars where both the UV and X-ray decline at similar ages \citep[][]{Shkolnik2014}. In comparing the fractional X-ray (coronal) and Ly$\alpha$ (chromospheric) flux to bolometric luminosity, \citet{Linsky2020}  observed that the relative emission from the chromosphere and corona show different distributions between M and K stars, potentially representing different heating mechanisms between the chromospheres and coronae of these stars. The corona can also heat the transition region through thermal conduction in a process known as back-heating \citep[e.g.,][]{Kopp1968, Athay1990}. Therefore, the FUV lines in this study that map the transition region may also experience a different evolution than the X-rays or other UV chromospheric lines.

\section{Discussion: What is causing the prolonged saturation period in K stars?}\label{sec:discussion}

\subsection{A Result of Different Initial Rotation Periods?}
Several studies \citep[e.g.][]{Tu2015, gondoin18, Magaudda2020} have shown that the initial rotation period of low-mass stars affects the evolution of the X-ray flux. Stars with faster initial rotation periods are saturated for longer and decline at a faster rate; conversely, stars with slower initial rotation periods decline in flux at younger ages, but decline at a lower rate. This trend most likely continues into the UV and accounts partially for the large scatter seen in \citet{richey-yowell2019} among K dwarfs. However, the rotation rates of our sample span a range of periods, yet we see constant flux for those at shorter or longer periods within the same age group. Therefore, it is unlikely that the prolonged period of saturated activity observed is due to a selection of only stars with faster original rotation periods.

\subsection{A Result of a Stalling in K Dwarf Rotational Spin-Down?}

Using K2 data from open clusters, \citet{Curtis2019, Curtis2020} show that current gyrochronological models fail to predict the rotation period of K stars beyond 650 Myr. Instead, the K stars appear to stall their spin-down, potentially maintaining their rotation period for over 1.3 Gyr before resuming to spin down. This effect is not seen in the same rotation data from \citet{Curtis2019} for G stars and it remains unknown if this occurs in M stars.

 \citet{Spada2020} were able to closely recreate these observations using a two-zone model for the star, where differential rotation between the convective envelope and the radiative core develops as magnetic wind braking reaches a maximum. The stalling may occur due to a balance between the spin-down torque from the magnetic wind braking and a spin-up torque from the transfer of angular momentum from the more quickly rotating radiative core. The redistribution of angular momentum is highly mass dependent, therefore the lower mass stars (i.e., K stars) are stalled for longer periods. \citet{Curtis2020}, however, do not rule out other theories such as reduced magnetic braking. If the core-envelope recoupling is the underlying method for the stalling, then this would imply that M dwarfs would not experience this effect, since stars become more fully convective with lower mass, and that there is perhaps a dominant relative size of the convection zone at which stalling is more pronounced. 

This phenomenon could explain the prolonged saturation period seen in the results presented above. If the K stars remain at a faster rotation period through a couple of Gyrs, than we would expect the evolution with age to follow a similar trend to that of the evolution of rotation period with age, maintaining a faster rotation period and thus a higher flux for longer periods of time. Similarly, flux as a function of rotation period would not be dissimilar from a case in which the spin-down of the star did not stall, which indeed is what we see when comparing the K stars with the M stars.

\section{Summary}\label{sec:conclusion}
In this paper, we present HST/COS UV spectra of 39 K stars at ages 40 Myr, 650 Myr, and $\approx$5 Gyr. Combining new observations of 28 K stars with 11 archived field-aged K dwarfs, we compared the evolutionary trends of K stars as a function of age, rotation, and Rossby number. 

Specifically, we analyze the evolution of the surface flux of the strong UV emission lines, i.e. C III, Si III, N V, C II, Si IV, C IV, He II, Mg II, as well as the FUV and NUV continua. We find that for each emission line and the pseudo-continua, the flux remains saturated between the 40 Myr and 650 Myr stars, before dropping off by up to two orders of magnitude. The UV evolution of K stars with age is thus distinct from that of early M stars, which begin to decline by a couple hundred Myr and show a clear decrease already by 650 Myr. However, when comparing the evolution of K and M star UV activity with rotation or Rossby number, this distinction goes away and the fits appear quite similar. 

This may imply that the rotational evolution of K star with age is dissimilar than that of M stars. This supports recent works by \citet{Curtis2019, Curtis2020}, which show that K dwarfs experience a period of time where their spin-down stalls, and the stars remain at a constant rotation period. This phenomenon is primarily thought to be caused by a core-envelope recoupling event and would weaken toward later type stars as the relative size of the convection zone grows, thus leading the early M dwarfs to decline in rotation rate and UV activity sooner than for K stars. 

K dwarfs have recently garnered attention for providing such a hospitable environment as to host potentially ``superhabitable'' planets \citep[e.g.][]{Cuntz2016, Heller2014, Arney2019}; yet, a prolonged period of saturated flux weakens this case. Further observations of open cluster members between 650 Myr -- 3 Gyr will be required to resolve at what age K dwarf rotational spin-down resumes. The delayed time evolution of K stars would increase the accumulated UV flux incident on their planets if the saturated period lasts for multiple Gyr. Additionally, a comparison of the flare evolution between these two types of stars will contribute to understanding the atmospheres of planets around these types of stars.

\vspace{0.5cm}
The authors would like to thank Dr. Jeff Linsky for a timely and insightful referee report. Support for this work was provided by NASA through grants numbered HST-GO-15091.001-A and HST-GO-15955.01 from the Space Telescope Science Institute, which is operated by AURA, Inc., under NASA contract NAS 5-26555. T.R.Y. would like to acknowledge additional support from the Future Investigators in NASA Earth and Space Exploration (FINESST) award 19-ASTRO20-0081 and thanks Aishwarya Iyer for useful MCMC discussions. This research has made use of the SIMBAD database, operated at CDS, Strasbourg, France. This work has made use of data from the European Space Agency (ESA) mission \textit{Gaia} (\url{https://www. cosmos.esa.int/gaia}), processed by the \textit{Gaia} Data Processing and Analysis Consortium (DPAC, \url{https://www.cosmos.esa.int/web/gaia/dpac/consortium})

\software{Astropy \, \citep{astropy:2018},\, Matplotlib \citep{matplotlib},\, Numpy \,\citep{numpy2}, \, Scipy \, \citep{scipy},\, emcee \citep[][]{Foreman-Mackey2013}, \, lightkurve \citep{Lightkurve}, \, stardate \citep[][]{Angus2019}}

\bibliography{bibliograph.bib}

\end{document}